\documentstyle[preprint,aps]{revtex}
\begin{document}
\draft
\preprint{}
\title{Exact Schwarzschild-Like Solution for Yang-Mills Theories}
\author{D. Singleton}
\address{Department of Physics, University of Virginia,
Charlottesville, VA 22901}
\date{\today}
\maketitle
\begin{abstract}
Drawing on the parallel between general relativity and Yang-Mills
theory we obtain an exact Schwarzschild-like solution
for SU(2) gauge fields coupled to a massless scalar field.
Pushing the analogy further we speculate
that this classical solution to the Yang-Mills equations shows
confinement in the same way that particles become confined
once they pass the event horizon of the Schwarzschild solution.
Two special cases of the solution are considered.
\end{abstract}
\newpage
\narrowtext
\section{Introduction}
Exact solutions to non-linear field theories are notoriously
difficult to find since there exists no general method for
discovering them. The usual approach is to make some
guess as to the form of the solution, and insert it into the
field equations to see if it solves them. The Schwarzschild
and the Kerr metrics are examples of such exact solutions
for Einstein's field equations. For Yang-Mills
theories there are also some known exact solutions such as
Coleman's plane wave solution \cite{coleman} and the Prasad-
Sommerfield solution \cite{prasad}. One possible avenue for
discovering new solutions for the Yang-Mills equations is to
exploit the long known parallel \cite{utiyama} that exists
between Yang-Mills gauge theories and general relativity, and
to see if some of the known solutions of general relativity
can be used as a guide for finding Yang-Mills solutions. Of
particular interest would be finding the Yang-Mills equivalent
of the Schwarzschild solution, since it exhibits a property
that has long been looked for in Yang-Mills theories, namely
confinement. Once a particle crosses the event horizon of the
Schwarzschild solution it becomes confined to the region inside
the event horizon.

In this note we give the exact, Schwarzschild-like solutions
for the Yang-Mills equations for an SU(2) gauge field coupled
to a massless scalar. It is argued that this classical solution
exhibits the property of confinement that has been looked for in
non-Abelian gauge theories.

\section{The Schwarzschild-Like Solution}

The model which we consider here is an SU(2) gauge field coupled
to a massless scalar triplet. The Lagrangian for this theory is
\begin{equation}
\label{lagran}
{\cal L} = -{1 \over 4} F^{\mu \nu a} F_{\mu \nu} ^a + {1 \over 2}
D^{\mu} ( \phi ^a ) D_{\mu} ( \phi ^a)
\end{equation}
where
\begin{equation}
F_{\mu \nu} ^a = \partial _{\mu} W_{\nu} ^a - \partial _{\nu}
W_{\mu} ^a + g \epsilon ^{abc} W_{\mu} ^b W_{\nu} ^c
\end{equation}
and
\begin{equation}
D_{\mu} \phi ^a = \partial _{\mu} \phi ^a + g \epsilon ^{abc}
W_{\mu} ^b \phi ^c
\end{equation}
We are interested in static solutions so all the time derivatives
will be zero. With this condition the gauge field equations
from Eq. (\ref{lagran}) are
\begin{equation}
\label{eqn1}
\partial _i F^{\mu i a} + g \epsilon ^{abc} W_i ^b F^{\mu i c} =
g \epsilon ^{abc} (D^{\mu} \phi ^b) \phi ^c
\end{equation}
For the scalar fields the equations are
\begin{equation}
\label{eqn2}
\partial _i ( D^i \phi ^a ) + g \epsilon^{abc} W _{\mu} ^b
( D ^{\mu} \phi ^c ) = 0
\end{equation}
Further assuming that the gauge fields and scalar fields are radial
we use the Wu-Yang ansatz \cite{wu}
\begin{eqnarray}
\label{ansatz}
W_i ^a &=& \epsilon_{aij} {r^j \over g r^2} [ 1 - K(r)] \nonumber \\
W_o ^a &=& {r^a \over g r^2} J(r) \nonumber \\
\phi ^a &=& {r^a \over g r^2} H(r)
\end{eqnarray}
Inserting this ansatz into the field equations of Eqs. (\ref{eqn1}),
(\ref{eqn2}) yields three coupled non-linear differential equations
\cite{zee}
\begin{eqnarray}
\label{nlpdd}
r^2 K'' &=& K (K^2 + H^2 - J^2 -1) \nonumber \\
r^2 J ''&=& 2JK^2 \nonumber \\
r^2 H'' &=& 2HK^2
\end{eqnarray}
The scalar field function, $H(r)$, and the time component of the
gauge field function, $J(r)$, enter the above equations in almost
the same way except for a difference in sign in the first equation.
The $W_0 ^a$ components act like an isotriplet scalar field with a
negative metric. The task of finding gauge and scalar fields
that solve Eqs. (\ref{eqn1}) and (\ref{eqn2}) thus simplifies
somewhat into the task of finding three functions, $K(r), J(r)$
and $H(r)$, which satisfy Eq. (\ref{nlpdd}). One such solution was
discovered by Prasad and Sommerfield \cite{prasad} in their
investigations of 't Hooft-Polyakov monopoles and Julia-Zee
dyons. Here we present another exact solution which was found
by using the connection between Yang-Mills theory
and general relativity.

Roughly speaking the objects in general relativity which
correspond to the gauge fields, $W_{\mu} ^a$,
are the connection coefficients, $\Gamma ^{\alpha}
_{\beta \gamma}$ (Ref. \cite{ramond}, which develops gravity as a
gauge theory, shows in what respects this last statement is a
bit of an oversimplification). Looking at some of the connection
coefficients of the Schwarzschild solution from general relativity
we find
\begin{eqnarray}
\Gamma ^t _{r t} &=& {K \over 2 r}{1 \over ( r - K )} \nonumber \\
\Gamma ^r _{r r} &=& -{K \over 2 r}{ 1 \over ( r - K )}
\end{eqnarray}
where $K = 2GM$. Using these connection coefficients as a guide,
and taking into account that there is an explicit $1 / r$
factor already in the ansatz of Eq. (\ref{ansatz}), we are led
to try the following solution
\begin{eqnarray}
\label{soln}
K(r) &=& {C r \over 1 - C r} \nonumber \\
J(r) &=& {B \over 1 - C r} \nonumber \\
H(r) &=& { A \over 1 - C r}
\end{eqnarray}
where $A, B$ and $C$ are arbitrary constants. It is a straightforward
exercise to insert these expressions for $K(r)$, $J(r)$ and $H(r)$ into
the coupled differential equations of Eq. (\ref{nlpdd}) and check that
they are solutions. The only constraint imposed is that $A^2 - B^2 = 1$,
so that the solution of Eq. (\ref{soln}) involves only two arbitrary
constants. Inserting $K(r), J(r)$ and $H(r)$ into the expressions
for the gauge and scalar fields of Eq. (\ref{ansatz}) it is seen
that both the gauge and scalar fields become infinite at the radius
\begin{equation}
r_0 = {1 \over C}
\end{equation}
Further, using these singular gauge potentials to calculate the
``electric'' and ``magnetic'' fields ($ E ^i _a = F^{i0} _a$
and $B ^ i _a = - {1 \over 2} \epsilon ^{ijk} F^{jk} _a$
respectively) it is seen that these fields are also infinite
at $r_0 = 1 / C$. Therefore a particle which carries an SU(2)
gauge charge becomes permanently confined if it crosses
into the region $r < r_0$. The non-Abelian gauge potentials
and the scalar fields of Eq. (\ref{ansatz}) also become singular at
$r= 0$, which is true as well for the Schwarzschild solution and for
the Coulomb potential of a point charge in classical electromagnetism.
The singularity of all these solutions at $r = 0$ are of the
same character in that they all imply a delta function
point ``charge'' sitting at the origin (where the charge of
general relativity is mass-energy, and the charge of our non-Abelian
model is SU(2) color charge). The singularity in our solution at
$r = 1/C$ is has an esstentially different character than the
Schwarzschild horizon in general relativity. The Schwarzschild
horizon is not a true singularity, but rather it is a coordinate
singularity which arises because of the choice of coordinates.
This can be seen when the Schwarzschild solution is given in
Kruskal coordinates which only have a singularity at $r=0$. For
our SU(2) solution the singularity at $r= 1/C$ can not be
transformed away by chosing a different coordinate system,
so it is a real singularity. Just as the singularity at the
origin can be taken to be a point source of SU(2) charge, so
the singularity at $r = 1/C$ can be taken to be a spherical
shell of SU(2) charge. This shell structure is a unique feature
of our solution, and it points to a possible connection with
the various phenomenological bag models of quark bound states.

There are two special cases which can be considered. First there
is the case where the time component of the gauge field equals
zero. This corresponds to taking $B= 0$ in Eq. (\ref{soln}). The
condition $A^2 - B^2 = 1$ then implies that $A = \pm 1$ so
that the solution becomes
\begin{eqnarray}
K(r) &=& {C r \over 1 - C r} \nonumber \\
H(r) &=& {\pm 1 \over 1 - C r}
\end{eqnarray}
There is also the unusual pure gauge case where there is no
scalar field. This corresponds to $H(r) = 0$ which implies
$A = 0$. The condition $A^2 - B^2 =1$ then requires that
$B = \pm i$ so that the solution becomes
\begin{eqnarray}
K(r) &=& {C r \over 1 - Cr} \nonumber \\
J(r) &=& {\pm i \over 1 - Cr}
\end{eqnarray}
It may seem strange to have a pure imaginary potential, however
it does solve the field equations. When these gauge fields are used
to calculate the energy in the fields one obtains a real, although
trivial answer ({\it i.e.} we will find that the field energy is zero).
Therefore one should be wary of the physical significance of this
pure gauge case.

The energy of the various gauge and scalar field configurations
of this Schwarzschild-like solution can be obtained by taking the
volume integral of the time-time component of the energy-momentum
tensor
\begin{eqnarray}
T ^{\mu \nu} &=& {2 \over \sqrt{ -g} }
{ \partial ( {\cal L} \sqrt{ -g} ) \over \partial g_{\mu \nu} }
\nonumber \\
&=& F^{\mu \rho a} F_{\rho} ^{\nu a} + D^{\mu} \phi ^a D^{\nu} \phi ^a
+ g ^{\mu \nu} \cal{ L }
\end{eqnarray}
The energy in the fields is then
\begin{eqnarray}
\label{mass}
E &=& \int T^{00} d^3 x \nonumber \\
&=& {4 \pi \over g^2} \int _{r_c} ^{\infty} \left( {K'} ^2
+ {(K ^2 - 1) ^2 \over 2 r^2}  + { J^2 K^2 \over r^2} +
{(r J' - J) ^2 \over 2 r^2} + {H^2 K^2 \over r^2} +
{(rH' - H) ^2 \over 2 r^2} \right) dr
\end{eqnarray}
Notice that the integral has been cut off from below at an
arbitrary distance $r_c$, which can be $> r_0$ or $< r_0$.
This was done to avoid the singularity at $r = 0$, since
integrating down to $r = 0$ would give an infinite field
energy in the same way that the Coulomb potential of
a point electric charge yields an infinite field energy
when integrated down to zero. An additional argument for
introducing the cutoff $r_c$ is the fact that our classical
solution does not exhibit asymptotic freedom. In this light
$r_c$ could be taken to delineate the boundary between the
region where our classical field solution dominates and the
region where the quantum effect of
asymptotic freedom dominates. Letting
the scalar field have a mass and a self coupling might smooth
out the behaviour of the fields at the origin, as is the case
with the 't Hooft-Polyakov \cite{thooft} monopole solution.
However when the scalar field is allowed to have a mass and
self coupling we can find no analytical solution, so numerical
solutions must be used. Inserting $K(r), J(r)$ and $H(r)$ into
Eq. (\ref{mass}) we find
\begin{eqnarray}
\label{mass2}
E &=& {2 \pi \over g^2} (A^2 + B^2 +1){(1 - 2 C r_c) \over
r_c (1 - C r_c) ^3} \nonumber \\
&=& {4 \pi A^2 \over g^2} {(1 - 2C r_c) \over r_c
(1 - C r_c) ^3}
\end{eqnarray}
where in the last expression the condition on the constants,
$A^2 - B^2 = 1$, has been used. In the special case where there
is only a scalar field and the space components of the gauge
fields the energy in the fields is given by the above expression
with $A^2 =1$ and $B^2 = 0$. For the pure gauge case $A^2 = 0$
and $B^2 = -1$ so that the energy of Eq. (\ref{mass2})
becomes zero. This together with the requirement that the
$W_0 ^a$ components of this solution are pure imaginary raises
doubts about the physical importance of this special case.

If this general Schwarzschild-like solution
is responsible for the confinement mechanism in
non-Abelian field theories, and if we discard the zero energy
pure gauge case, then it is necessary for the Lagrangian to
always include scalar fields in order for an acceptable solution to
exist. Under these assumptions scalars fields become crucial to
the confinement mechanism. This is in contrast to the conventional
ideas about confinement, where scalar fields do not play a role.

One possible test for the physical importance of this solution to
strong interaction physics would be to see if it could be used
to calculate the consistiuent masses of the light quarks in their
various bound states (mesons and baryons). For the light quark
bound states ({\it e.g.} protons, pions) most of the mass is believed
to reside in the gluon fields rather than in the current quark
masses. Before any numerical results could be extracted it would
first be necessary to specify the arbitrary constant $C$. The
equivalent object in the Schwarzschild solution of general relativity
is $1/(2GM)$. Using the analogy between
Yang-Mills and general relativity
it can be argued that $C$ should be related to the strength of the
interaction and the magnitude of the ``color'' charge carried by the
quark. Also one should be able to get an experimental estimate for
this constant since $1/C$ should be roughly related to the
radius of the bound state. Second, our present solution is for
an SU(2) gauge theory while QCD is formulated in terms of the
SU(3) gauge group. Thus one would need to generalize the present
solution to SU(3) or if possible to SU(N).
This should be possible by embedding the SU(2) solution in the
higher rank gauge group. Finally the most serious obstacle to using
this Schwarzschild-like solution to calculate QCD bound states is
that the field equations are highly non-linear so that the
superposition of two solutions will not necessarily be a solution.
However the present solution might provide a framework for a numerical
calculation, or for some approximate phenomenological development
for obtaining rough estimates for the consistiuent quark masses.

Finally, just like the Prasad-Sommerfield solution \cite{prasad}, our
solution can be seen to carry a topological magnetic charge when the
electromagnetic field is embedded into the SU(2) theory via 't Hooft's
\cite{thooft} generalized, gauge invariant, electromagnetic field
strength tensor
\begin{equation}
\label{maxwell}
{\cal F}_{\mu \nu} = \partial_{\mu} (\hat{\phi} ^a W_{\nu}^a) -
\partial_{\nu} (\hat{\phi} ^a W_{\mu}^a) - {1 \over g} \epsilon^{abc}
\hat{\phi} ^a (\partial_{\mu} \hat{\phi} ^b) (\partial_{\nu} \hat{\phi} ^c)
\end{equation}
where $\hat{\phi} ^a = \phi ^a (\phi ^b \phi ^b)^{-1/2}$. Using this
generalized electromagnetic field strength tensor the magnetic field
of our solution is
\begin{equation}
{\cal B}_i = {1 \over 2} \epsilon_{ijk} {\cal F} _{jk} = -{r_i \over g r^3}
\end{equation}
which is the magnetic field of a point monopole of strength ${-4 \pi \over g}$
at the origin. Looking in detail at Eq. (\ref{maxwell}) it can be seen that
only the last term contributes to the magnetic charge of the solution. The
fact that the magnetic charge comes entirely from the scalar field ,
in this gauge, has been
explained in terms of the topology of the scalar field \cite{arafune}. The
electric field of our solution is different for that of the
Prasad-Sommerfield solution. Using the generalized electromagnetic
field strength tensor we find that the electric field of our solution is
\begin{equation}
\label{efield}
{\cal E}_i = -{\cal F} _{0i} = {r_i \over r} {d \over dr}
\left( {J(r) \over r} \right) = {B (2Cr - 1) r_i \over r^3 (1 - Cr)^2}
\end{equation}
As $r \rightarrow \infty$ this electric field falls off like
$1 / r^3$, unlike the Prasad-Sommerfield solution which has a
$1 / r^2$ behaviour for large $r$. In the Prasad-Sommerfield
case the electric field at large $r$ indicated the presence of
some net charge which was distributed in a cloud around the
origin. For our solution the behaviour of the electric field
at large $r$ indicates that while there is no net charge, there
is some distribution of charge which has nonzero higher order
moments (the $1/ r^3$ behaviour of the electric field is like
that of a dipole charge distribution, although the electric
field of Eq. (\ref{efield}) is definitely not that of a point
dipole). The charge density of our solution can easily be found
by applying $\rho(r) = \nabla \cdot {\bf {\cal E}}$ to the electric
field of Eq. (\ref{efield}). Notice that while the topological
magentic charge associated with our solution is similiar to that
of the Prasad-Sommerfield solution, the non-topological electric
charge is significantly different.

\section{Conclusions}

Drawing on the parallels between general
relativity and Yang-Mills theory we have discovered an exact
Schwarzschild-like solution for an SU(2) gauge field coupled
to a massless scalar field. The general solution involved
both the time and space components of the gauge fields as well
as scalar fields. There were two special cases which occured :
First the time component of the gauge fields could be zero leaving
only the scalar fields and space component of the gauge field;
second there was the pure gauge solution, where the scalar fields
were absent, leaving only the time and space components
of the gauge fields. These classical solutions exhibited a form
of confinement. Any particle which carries an SU(2) gauge charge
and enters the region $r < r_0 = 1 / C$ would no longer be able
to leave this region. This is analogous to what happens
with the Schwarzschild solution in general relativity, where once
a particle passes the event horizon it is permanently confined.
This is still a long way from demonstrating confinement for QCD,
since our entire argument has been for classical Yang-Mills fields,
and the ansatz that we adopted is specific to the SU(2) group,
depending on the fact that the number of SU(2) generators and the
number of spatial dimensions are the same. Still it is encouraging
that at the classical level an analytical solution, which seems
to exhibit confinement, can be found for a non-Abelian gauge
theory.

There are several open questions which the present letter does not
discuss. First, are there other known solutions of general relativity
which can be translated into Yang-Mills gauge theories ? One interesting
possibility would be to see if the axially symmetric, rotating
mass solution (the Kerr metric) could be used to discover an analogous
Yang-Mills solution. Second, just as a Schwarzschild black hole is
thought to emit Hawking radiation, the Schwarzschild-like solution
presented here might be conjectured to exhibit a similiar effect
with respect to virtual particle pairs which carried the non-Abelian
gauge charge. Before such a conjecture can be checked the present
classical field solution would have to be quantized.

\section{Acknowledgements} The author wishes to acknowledge the
help and suggestions of James Singleton and Justin O' Neill. For
the partial financial support of this work the author thanks
PVPC and Don Jones.

\end{document}